\documentclass[a4paper,11pt]{article}
\usepackage{cite}
\usepackage{amssymb}
\usepackage{graphicx}
\usepackage[unicode=true,pdfusetitle,bookmarks=true,bookmarksnumbered=false,bookmarksopen=false,
breaklinks=false,pdfborder={0 0
1},backref=false,colorlinks=false]{hyperref}
\usepackage{mathtools}
\usepackage{slashed}
\begin{document}
\def\boxit#1{\vcenter{\hrule\hbox{\vrule\kern8pt
      \vbox{\kern8pt#1\kern8pt}\kern8pt\vrule}\hrule}}
\def\Boxed#1{\boxit{\hbox{$\displaystyle{#1}$}}} 
\def\sqr#1#2{{\vcenter{\vbox{\hrule height.#2pt
        \hbox{\vrule width.#2pt height#1pt \kern#1pt
          \vrule width.#2pt}
        \hrule height.#2pt}}}}
\def\square{\mathchoice\sqr34\sqr34\sqr{2.1}3\sqr{1.5}3}
\def\Square{\mathchoice\sqr67\sqr67\sqr{5.1}3\sqr{1.5}3}
\title{\bf QED Treatment of Linear Elastic Waves in Asymmetric Environments}
\author{Maysam Yousefian\thanks{E-mail: M\_Yousefian@sbu.ac.ir}\ \ and Mehrdad Farhoudi\thanks{E-mail:
 m-farhoudi@sbu.ac.ir}\\
 {\small Department of Physics, Shahid Beheshti University G.C.,
         Evin, Tehran 19839, Iran}}
\date{\small May 16, 2021}
\maketitle
\begin{abstract}
\noindent
 Considering the importance of correctly understanding
the dynamics of microstructure materials for their applications in
related technologies, by eliminating the shortcomings and some
overlooked physical concepts in the existing asymmetric elastic
theories, we have presented an asymmetric elastodynamic model
based on a $U(1)$ gauge theory with quantum electrodynamics ({\bf
QED}) structure. Accordingly, we have shown that there is a
correspondence between an elastic theory, which can explain the
behavior of elastic waves within an asymmetric elastic medium, and
QED. More specific, we have indicated that the corresponding
elastic wave equations are somehow analogous to QED ones. In this
regard, by adding vibrational degrees of freedom and introducing a
gauge property of the waves of displacement for the waves of
rotation, we have generalized and modified the related Cosserat
theory ({\bf CT}) for an elastic environment. Thus on macro
scales, the elastic waves can possess the QED treatment. This
analogy provides a new paradigm of fermions and bosons. Also, from
experimental point of view, we have shown that the behavior of
elastic waves in a granular medium is equivalent to behavior of
light in dispersive media, which can be explained using QED.
Hence, contrary to the Cosserat and discrete models, this amended
CT has qualitatively been indicated to be consistent with the
corresponding empirical observations.
\end{abstract}
\bigskip
\section{Introduction}
\indent

There is no doubt that one of the most important and essential
discussions in technologies, which are based on polymers,
crystals, granular and cellular materials, are the understanding
of internal dynamics among the structures of these materials that
have microstructures. As recently, important and significant
researches have been assigned on these topics. For example, a lot
of articles have been presented on wave propagation in composite
materials, see e.g.,
Refs.~\cite{Karami--2019,Karami--2019-2,Furjan--2020,Asghar--2020,Khadimallah--2020,Furjan--2021},
and on the dynamic properties of nanocomposite materials, see
e.g.,
Refs.~\cite{Batou---2019,Bouanati--2019,Furjan---2020,Karami---2019,Bendenia---2020,Heidari---2021,Arshid---2021}.
In this regard, it is known that discrete techniques can supply a
more accurate explanation of discontinuous materials. However,
these techniques are complex and necessitate accuracy in modeling
of the interface. For instance, inappropriate selection of
interface elements in many cases leads to realms of
unrealistically high stress gradients and incorrect results. In
addition, in many applications, the definition of the input model
becomes impractical as the number of joints gets
large~\cite{Riahi}. Such issues can be considered as a challenge
for discrete models.

In the equivalent continuum techniques, the properties of the
interface are mixed with the properties of the intact matrix, and
discrete materials are replaced by homogeneous
continuums~\cite{Riahi}. The equivalent continuum techniques, that
are based on the classical continuum theory, have the challenge of
not taking into account the effect of the internal length scale,
which is an intrinsic characteristic of layered materials.
Consequently, these techniques are suitable to cases where no slip
occurs at the interface of the layers and/or when the internal
length scale of the problem is negligible compared to the
dimensions of the engineering structure~\cite{Riahi}.

Given the important challenges of current discrete and continuous
elastic models, for a better description of dynamics of elastic
environments, one needs a model without the shortcomings of the
current models. Accordingly, from theoretical and experimental
points of view, via generalization and modification of some
existing elastic models, we have investigated relevance between
such an elastic theory (which should explain the behavior of
elastic waves within an asymmetric elastic medium) and a $U(1)$
gauge theory (e.g. QED). In particular, we have focused on the
equations of motion of elastic waves in asymmetric environments
and the corresponding QED ones. In this respect, in addition to
the advantages of those continuous elastic models, according to
the quantum aspects (e.g. loop corrections), a $U(1)$ gauge theory
(e.g. QED), as a continuous model, has other properties that can
compensate for the shortcomings of discrete models. Indeed, due to
the phenomenon of loop corrections and the existence of cut-off
energy in QED, such a relevance can be important when it is
employed to describe the effect of internal length scale of
discontinuous materials. In this process, we have been looking
forward that one would be able to take advantage of continuum
techniques without the challenge in describing discontinuous
materials.

Previous attempts had been to achieve an elastic model with
respect to electromagnetic theory. However, in the scope of
classical elastic theory\rlap,\footnote{In the classical elastic
theory, the stress tensor has normally been considered to be
symmetric.}\
 the similarity between the elastic and electromagnetic waves is
incomplete. For example, in
Refs.~\cite{Kelvin,Dmitriyev92,Dmitriyev98}, an analogy between
the electromagnetic field and a fluid, which includes a large
number of vortex filaments, has been investigated. In these works,
the pressure of fluid, the fluid velocity and the density of the
turbulence energy respectively are as the electrostatic potential,
the magnetic vector potential and the electromotive force, but
charge has no independent nature. Also in
Refs.~\cite{Dmitriyev01,Yoshida}, it has been claimed that the
electromagnetic equations fit the structure of a linear elastic
continuum that is hard to compression though liable to shear
deformation. Nevertheless, in these two works, charge still has
no~independent nature because the microstructures have~not been
considered for the elastic environment. Whereas a physical elastic
environment is composed of microstructures.

On the other hand, in the
CT~\cite{Cosserat,ref28,Toupin,Vardoulakis}, by generalizing the
elastic theory and assuming the presence of microstructures in an
elastic environment, the stress tensor has been employed
asymmetrically. However, still there are some overlooked physical
concepts in these theories. For instance, any microstructure can
have a vibrational degree of freedom, which has~not been talked
about in these theories. Whereas, the waves of displacement can
play a role in transmission of the waves of rotation as a gauge
property\rlap,\footnote{The displacement waves play the role of a
gauge field for the rotation waves and can, in a way, change the
transmission speed or momentum of the rotation waves. This
behavior is similar to the behavior of a photon relative to an
electron in QED.}\
 but there is no mention of the gauge property
of the waves of displacement for the waves of rotation of the
microstructures.

In this regard, to achieve an asymmetric elastodynamic model based
on the QED as a $U(1)$ gauge theory with quantum aspects, we have
generalized and then have modified the CT. During this process,
first by adding vibrational degrees of freedom for microstructures
in the CT and using some mathematical methods, we have achieved a
covariant form of the CT. Then, while using the gauge property of
the waves of displacement and also modeling fermions, by means of
the Cosserat elasticity\rlap,\footnote{In principle, the idea of
modeling fermions by means of the Cosserat elasticity is a
challenging task, though there are~not that many publications on
the subject. Somehow this approach has never really taken off.}\
 we have modified the CT in a way that the resulted asymmetric elastodynamic
model being fully comparable to QED. Finally, we have made a
comparison between the experimental observations and the
predictions of the proposed asymmetric elastodynamic model.

\section{The Cosserat Theory with Minkowski Metric}
\indent

In the absence of any external force and torque, the corresponding
dynamical equations for any homogeneous and isotropic linear
asymmetric elastic environment, with the Minkowski metric, can be
written~\cite{ref29} as\footnote{In Ref.~\cite{ref29}, these
equations have been written in the Euclidean three-dimensional
space, however, we have rewritten those based on the Minkowski
metric with signature $+2$. Also, we indicate the partial
derivatives with commas and use the Einstein summation
convention.}
\begin{equation}\label{eq1}
\rho\, u^{i,0}{}_{0}+\sigma^{ji}{}_{,j}=0,
\end{equation}
\begin{equation}\label{eq1-1}
\Im\,\varphi^{i,0}{}_{0}+\epsilon^{ijk}\sigma_{jk}+\mathfrak{m}^{ji}{}_{,j}=0,
\end{equation}
where $ \rho $ is constant average mass density of environment, $
\Im $ is the density of moment of inertia, $ \sigma^{ij} $ is the
stress tensor, $ \mathfrak{m}^{ij} $ is the couple-stress tensor,
$ u^{i} $ and $ \varphi^{i} $ respectively are the displacement
vector and the rotation pseudovector of any point of environment,
and $\epsilon^{ijk}$ is the Levi-Civita alternating symbol. As the
displacement and rotation satisfy wave equations and are defined
over the entire elastic environment, we also refer to these as
waves or fields. The Latin lowercase letters run from one to
three, and the unit of speed has been selected to be one.

As Eqs. (\ref{eq1}) and (\ref{eq1-1}) contain four unknown
parameters, in order to proceed, we appeal to the definitions of
deformation tensors as~\cite{ref29}\footnote{$\Upsilon_{ij} $ is
the deformation gradient tensor.}
\begin{equation}\label{eq1-2-a}
\Upsilon_{ij}\equiv u_{j,i}-\epsilon_{ijk}\,\varphi^{k}
\qquad\qquad{\rm and}\qquad\qquad \chi_{ij}\equiv\varphi_{j,i}.
\end{equation}
In addition, when an environment behaves linearly, it has been
shown~\cite{ref29} that one obtains
\begin{equation}\label{eq1-2}
\sigma^{ij}=(\alpha+\mu)\Upsilon^{ij}+(\mu-\alpha)\Upsilon^{ji}+\lambda\delta^{ij}\Upsilon^{k}{}_{k},
\end{equation}
\begin{equation}\label{eq1-3}
\mathfrak{m}^{ij}=(\gamma+\varepsilon)\chi^{ij}+(\gamma-\varepsilon)\chi^{ji}+\beta\delta^{ij}\chi^{k}{}_{k},
\end{equation}
where $\delta^{ij}$ is the Kronecker delta.  The six parameters
$\alpha$, $\mu$, $\gamma$, $\varepsilon $, $\lambda$ and $\beta$
are the characteristic coefficients of an elastic medium. These
coefficients are measured in an adiabatic mode and are constants
in a homogeneous environment, in a way that, the first four ones
have positive values and the last two ones are required to satisfy
the inequalities $3\lambda +2 \mu> 0$ and $3\beta + 2 \gamma>
0$~\cite{ref29}. Accordingly, substituting
definitions~(\ref{eq1-2}) and (\ref{eq1-3}) with relation
(\ref{eq1-2-a}) into Eqs.~(\ref{eq1}) and (\ref{eq1-1}), these
equations read\footnote{In Ref.~\cite{ref29}, the term
$4\alpha\varphi^{i}$ in Eq.~(\ref{eq1-5}) is missing, and it has
been stated that total initial conditions have been marked with a
symbol therein. However, an initial condition is obviously
something different from a term in field equations, besides the
mentioned term is a field, which its existence changes the nature
of waves of rotation to a massive wave (as it will be described
below). Also, field equations have to be the same independently of
any initial conditions, hence for simplicity, we consider initial
conditions to be zero.}
\begin{equation}\label{eq1-4}
\rho\,
u^{i,0}{}_{0}+(\alpha+\mu)u^{i,j}{}_{j}+(\mu-\alpha+\lambda)u^{j}{}_{,j}{}^{i}+2\alpha\epsilon^{ijk}\varphi_{j,k}=0,
\end{equation}
\begin{equation}\label{eq1-5}
        \Im\,\varphi^{i,0}{}_{0}+(\gamma+\varepsilon)\varphi^{i,j}{}_{j}
        +(\gamma-\varepsilon+\beta) \varphi^{j}{}_{,j}{}^{i} -
        4\alpha\varphi^{i}-2\alpha\epsilon^{ijk}u_{j,k}=0.
\end{equation}

\section{Four-Dimensional Representation}
\indent

To achieve a `relativistic' appearance, we furthermore extend the
above obtained equations to four-dimensional spacetimes.

For this purpose, first to introduce a zeroth component for the
displacement, we utilize a scalar field in analogy with the issue
of longitudinal waves. In there, a scalar field (as a kinetic
potential) has been used to explain some parameters\footnote{In
linear acoustics, a real potential has been presented in a way
that the turbulence of pressure and mass density, and speed can be
provided in terms of it~\cite{Stanzial}.}\
 including the
condensation\rlap,\footnote{The ratio of difference between the
instantaneous and the equilibrium mass densities to the
equilibrium mass density of a medium, at a point, is called
condensation~\cite{Kinsler-1999}.}\
 which is defined as minus the time derivative of
it multiplied by the inverse square of the longitudinal component
of the speed of wave, $ v_{_{\rm{L}}} $, \cite{Stanzial}. Also, in
the linear acoustics, it has been shown~\cite{Kinsler-1999} that
the condensation is equal to minus the divergence of the
displacement in the Euclidean space. Thus, for such a scalar
field/kinetic potential, say $\phi $, one obtains
\begin{equation}\label{cont eq}
\phi_{,0}-v_{_{\rm{L}}}^{2}u^{i}{}_{,i}=0.
\end{equation}
Incidentally, the longitudinal component of the speed of wave is
related to the elastic environment characteristic coefficients
(the Lam\'{e} constants) as $ v_{_{\rm{L}}}^{2} \equiv (\lambda +
2 \mu) / \rho $~\cite{Harris2004}. With these respects, we rewrite
relation (\ref{cont eq}) as
\begin{equation}\label{continuity equation}
u^{\hat{\alpha}}{}_{,\hat{\alpha}} =0
\end{equation}
with metric $g^{\hat{\alpha}\hat{\beta}}\equiv
(-v_{_{\rm{L}}}^{-2},1,1,1)$ for hat-letters, where the kinetic
potential has been assumed as zeroth component of the
displacement, $\phi\equiv u_{0}$, and the Greek lowercase letters
run from zero to three.

\section{Multi-Metric Instead of Multi-Speed}
\indent

For simplicity of the appearances and also for clarity of the
resulted equations, to raise the covariant components of any
tensor (i.e., to get any covariant components from the
corresponding contravariant ones), we manifestly define five
different diagonal metrics specified with normal-letters,
hat-letters, bar-letters, check-letters and tilde-letters through
the work. We already have defined the metric with hat-letters, the
metric with the normal-letters is $g^{\alpha\beta}=(-1,1,1,1)$,
and the remaining ones will be defined later. Indeed, the need for
these five metrics is due to the fact that there are six
parameters characterizing the elastic continuum. These mentioned
metrics are defined in a way that only differ with each other in
their `timelike' $00$ components. These components are related to
different speeds (like $g^{\hat{0}\hat{0}}$), and actually, we
prefer to somehow specify those rather than assuming all of those
to be one (as in the usual natural units and/or in the case with
the normal-letters). Of course, we assume that, for all of these
different kinds of letters, the corresponding covariant components
of any tensor (obviously, except the corresponding metrics
themselves) being the same, say for example
$T_{\alpha}=T_{\hat{\alpha}}=T_{\bar{\alpha}}=T_{\check{\alpha}}=T_{\tilde{\alpha}}$.
Hence, as desired, this restriction makes only the contravariant
zeroth-components being different, e.g., $T^{\hat 0}=v^{-2}_L
T^0$. Nevertheless, for symmetry of appearances of relations,
wherein the summation rule is used, we specify the notation of any
covariant component with the same notation of its corresponding
contravariant one, e.g. as done in relation~(\ref{continuity
equation}).

With known definitions
\begin{equation}\label{eq1-14-1-1-2aaa-x}
\varphi^{ij}\equiv\epsilon^{ijk}\varphi_{k}\qquad\qquad {\rm and}
\qquad\qquad F^{\alpha\beta}\equiv
u^{\alpha,\beta}-u^{\beta,\alpha},
\end{equation}
while using relation \eqref{continuity equation}, we can rewrite
Eq.~\eqref{eq1-4} as
\begin{equation}\label{eq1-4-x}
(\alpha+\mu)
F^{i\bar{\alpha}}{}_{,\bar{\alpha}}-2\alpha\varphi^{ij}{}_{,j}=0
\end{equation}
with metric
$g^{\bar{\alpha}\bar{\beta}}\equiv(-v_{_{\rm{T}}}^{-2},1,1,1)$ for
bar-letters, where $v_{_{\rm{T}}}^2\equiv(\alpha+\mu)/\rho $ is
the transverse component of the speed of wave for a
divergence-free wave of displacement.

\section{Generalization of CT via New Degrees of Freedom}
\indent

We secondly need to introduce another three components for the
existing antisymmetric tensor of the wave of rotation $
\varphi^{ij}$, which we indicate those as $\varphi^{0i}$'s
components. For this purpose, as the moment of inertia density is
proportional to the mass density of environment, thus varying it
will cause that to change. The simplest way to explain variation
of moment of inertia density is to assume the microstructures of
medium as two same balls, each with mass $m$, connected by a
massless spring, i.e. a three-dimensional isotropic harmonic
oscillator where its symmetry group is isomorphic to $SO(4)$. To
show this, let us write the energy of these mentioned two balls as
the energy of another two balls, each also with mass $m$ in a
four-dimensional Euclidean space, connected by a massless rod of
length $2R$, which are rotating around their center of mass. That
is, a four-dimensional linear rigid rotor with total energy $E$
and angular momentum $l_{(4)} $ of each ball, where
$l_{(4)}^2\equiv l^{AB}l_{AB}/2 $ and the Latin uppercase letters
run from one to four. Then, one can easily show that
\begin{equation}\label{lfour}
\frac{l_{(4)}^2}{2mR^2}=\frac{m\dot{r}^2}{2}+\dfrac{Er^2}{R^2}+\dfrac{l^2}{2mr^2},
\end{equation}
where $ l$ and $ r$ respectively are the three-dimensional angular
momentum of each ball (with $l^2=l^{ij}l_{ij}/2 $) and the
distance between two balls in $3$-dimensions. To prove
relation~(\ref{lfour}), note that $x^A x_A=R^2 $ and $
\dot{x}^A\dot{x}_A=2E/m $ and $l^{AB}\equiv
mx^A\dot{x}^B-mx^B\dot{x}^A$. This relation indicates that the
symmetry group of the isotropic harmonic oscillator in
$3$-dimensions is isomorphic to $ SO(4)$.

In this way, we assume the microstructures of a medium being as
four-dimensional linear rigid rotors with four-dimensional angular
rotation $\varphi^{\alpha\beta}$, which $ \varphi^{0i} $'s (as
vibration and relative velocity of parts of each microstructure)
are related to $\varphi^{4i} $'s (as rotation in fourth dimension)
via the Wick rotation. On the other hand, by selecting an
appropriate gauge, we choose a gauge fixing in such a way that the
displacement and rotation wave equations of motion to be
covariant, and in general, the corresponding velocities of waves,
masses and coupling (to the displacement field) of $ \varphi^{0i}$
being different from those of $ \varphi^{ij}$. In this respect, we
assume the gauge
\begin{equation}\label{gauge-fixing}
\varphi^{\check{\gamma}\check{\alpha}}{}_{,\check{\gamma}}{}^{\check{\beta}}
-\varphi^{\check{\gamma}\check{\beta}}{}_{,\check{\gamma}}{}^{\check{\alpha}}
=0
\end{equation}
with metric $
g^{\check{\alpha}\check{\beta}}=(-v^{-2}_{\rm{V}},1,1,1) $ for
check-letters, where $v^{2}_{\rm{V}}\equiv\left(
\gamma+\varepsilon\right)/\Im $ is the speed of wave for
$\varphi^{0i} $ waves of vibration. Then, we plausibly generalize
Eq.~\eqref{eq1-5} as
\begin{equation}\label{rotation-eq-i}
        \left( 2\gamma+\beta\right)\left(
        \varphi^{\tilde{\alpha}\tilde{\beta}}{}_{,\tilde{\gamma}}{}^{\tilde{\gamma}}\right.
         \left.
        +\varphi^{\tilde{\gamma}\tilde{\alpha}}{}_{,\tilde{\gamma}}{}^{\tilde{\beta}}
        -\varphi^{\tilde{\gamma}\tilde{\beta}}{}_{,\tilde{\gamma}}{}^{\tilde{\alpha}}\right)
        -4\alpha\varphi^{\tilde{\alpha}\tilde{\beta}}-2\alpha
        F^{\bar{\alpha}\bar{\beta}}=0
\end{equation}
with metric $
g^{\tilde{\alpha}\tilde{\beta}}=(-v^{-2}_{\rm{R}},1,1,1) $ for
tilde-letters, where $v^{2}_{\rm{R}}\equiv\left(
2\gamma+\beta\right)/\Im $ is the speed of wave for $\varphi^{ij}
$ waves of rotation. In Eq.~\eqref{rotation-eq-i}, the part
\begin{equation}\label{rotation-eq-ij}
        \left( 2\gamma+\beta\right)\left(
        \varphi^{ij}{}_{,\tilde{\gamma}}{}^{\tilde{\gamma}}
        +\varphi^{\tilde{\gamma}i}{}_{,\tilde{\gamma}}{}^{j}-\varphi^{\tilde{\gamma}j}{}_{,\tilde{\gamma}}{}^{i}\right)
        -4\alpha\varphi^{ij}-2\alpha F^{ij}=0,
\end{equation}
using gauge \eqref{gauge-fixing}, is just Eq.~\eqref{eq1-5}, and
the extra part
\begin{equation}\label{rotation-eq-i0}
        \left( 2\gamma+\beta\right)\left(
        \varphi^{\tilde{0}i}{}_{,j}{}^{j}
        +\varphi^{j\tilde{0}}{}_{,j}{}^{i}
        -\varphi^{ji}{}_{,j}{}^{\tilde{0}}\right)
        -4\alpha\varphi^{\tilde{0}i}-2\alpha F^{\bar{0}i}=0,
\end{equation}
again due to gauge \eqref{gauge-fixing}, is the wave equation of
vibration $\varphi^{0i} $, i.e.
\begin{equation}\label{rotation-eq-i0-00}
\left( \gamma+\varepsilon\right)
\varphi^{\check{0}i}{}_{,\check{\gamma}}{}^{\check{\gamma}}
 -4\alpha\varphi^{\tilde{0}i}-2\alpha
 F^{\bar{0}i}=0.
\end{equation}

Eq.~\eqref{rotation-eq-i} is invariant under a gauge
transformation with respect to an arbitrary gauge field, say
$\zeta_{\alpha} $, as
\begin{equation}\label{gauge-trans-1}
\begin{split}
& \varphi^{\tilde{\alpha}\tilde{\beta}}\rightarrow
\varphi^{\tilde{\alpha}\tilde{\beta}}-\left(
\zeta^{\tilde{\alpha},\tilde{\beta}}-\zeta^{\tilde{\beta},\tilde{\alpha}}\right)
\\ &F^{ij}\rightarrow F^{ij}+2\left(
\zeta^{i,j}-\zeta^{j,i}\right) \\ &F^{i\bar{0}}\rightarrow
F^{i\bar{0}}+2\frac{v^{2}_{\rm{T}}}{v^2_{\rm{R}}}\left(
\zeta^{i,\bar{0}}-\zeta^{\bar{0},i}\right).
\end{split}
\end{equation}
This gauge gives us a cue to manage Eq.~\eqref{eq1-4-x} being also
invariant under a gauge transformation with respect to another
arbitrary gauge field, say $\xi_{\alpha} $, as
\begin{equation}\label{gauge-trans-2}
\begin{split}
& \varphi^{\tilde{\alpha}\tilde{\beta}}\rightarrow
\varphi^{\tilde{\alpha}\tilde{\beta}}+\xi^{\tilde{\alpha},\tilde{\beta}}-\xi^{\tilde{\beta},\tilde{\alpha}}
\\ &F^{ij}\rightarrow F^{ij}+\frac{2\alpha}{\alpha+\mu}\left(
\xi^{i,j}-\xi^{j,i}\right) \\ &F^{i\bar{0}}\rightarrow
F^{i\bar{0}}+\frac{2\alpha v^{2}_{\rm{T}}}{(\alpha+\mu)
v^2_{\rm{R}}}\left( \xi^{i,\bar{0}}-\xi^{\bar{0},i}\right),
\end{split}
\end{equation}
provided that we generalize Eq. \eqref{eq1-4-x} in a more general
form
\begin{equation}\label{eq1-4-xx}
(\alpha+\mu)
F^{i\bar{\alpha}}{}_{,\bar{\alpha}}-2\alpha\varphi^{i\tilde{\alpha}}{}_{,\tilde{\alpha}}=0,
\end{equation}
however with $\varphi^{0i}=0 $ in order to reproduce
Eq.~\eqref{eq1-4-x}. The generalization of adding such a degree of
freedom (i.e., $\varphi^{0i}$) is physically meaningful. The
reason is that, $ F^{0i} $ is composed of the velocity of
microstructures and gradient of the kinetic potential (that, in
turn, is related to change of the density of microstructures). On
the other hand, the vibration $\varphi^{0i} $ has also two
features, namely the relative velocity of parts of each
microstructure and alteration in the volume (and hence the
density) of each one. Thus, $ F^{0i} $ and $ \varphi^{0i} $ have a
similar role, which is missing in Eq.~\eqref{eq1-4-x}.

Now, performing the divergence of Eq. (\ref{eq1-4-xx}) yields
\begin{equation}\label{eq1-4000}
(\alpha+\mu) F^{i\bar{0}}{}_{,\bar{0}i}
-2\alpha\varphi^{i\tilde{0}}{}_{,\tilde{0} i}=0.
\end{equation}
Integrating Eq. (\ref{eq1-4000}) over time while considering the
zeroth component of the displacement as a wave, it
gives\footnote{Eq.~(\ref{eq1-40000}), without the second
interaction term while using relation (\ref{continuity equation}),
explains the kinetic potential as a wave~\cite{Stanzial}. }
\begin{equation}\label{eq1-40000}
(\alpha+\mu)
F^{\bar{0}\bar{\alpha}}{}_{,\bar{\alpha}}-2\alpha\varphi^{\tilde{0}\tilde{\alpha}}{}_{,\tilde{\alpha}}=0.
\end{equation}
Combining Eqs. \eqref{eq1-4-xx} and \eqref{eq1-40000}, we finally
end up with a generalized form of Eq.~\eqref{eq1-4-x} as
\begin{equation}\label{Neq1-40000}
(\alpha+\mu)
F^{\bar{\beta}\bar{\alpha}}{}_{,\bar{\alpha}}-2\alpha\varphi^{\tilde{\beta}\tilde{\alpha}}{}_{,\tilde{\alpha}}=0.
\end{equation}

Now that we have obtained the necessary equations in the covariant
form, it would be instructive to use the exterior algebra
notations\rlap,\footnote{See, e.g., Ref.~\cite{Strauman}.}\
 in order to concisely rewrite those. Indeed, it gives us a
better and simpler sight to investigate the correspondence between
the QED equations and the elastic wave ones. In this regard, we
respectively rewrite gauge conditions \eqref{continuity equation}
and \eqref{gauge-fixing}, and Eqs.~\eqref{rotation-eq-i} and
\eqref{Neq1-40000} as
\begin{equation}\label{dis-eq-xxx}
\hat{\delta}\hat{ \mathbf{u}}=0\quad\qquad {\rm and}\quad\qquad
\check{d} \check{\delta}\check{\boldsymbol\varphi} =0,
\end{equation}
\begin{equation}\label{dis-eq-x}
\left( 2\gamma+\beta\right)
\tilde{\delta}\tilde{d}\tilde{\boldsymbol\varphi}-4\alpha\tilde{\boldsymbol\varphi}-2\alpha\bar{d}\bar{
\mathbf{u}}=0,
\end{equation}
\begin{equation}\label{dis-eq}
(\alpha+\mu) \bar{\delta}\bar{d}\bar{
\mathbf{u}}-2\alpha\tilde{\delta}\tilde{\boldsymbol\varphi}=0,
\end{equation}
where $ \mathbf{u}$ and $\boldsymbol\varphi$ are one-form and
two-form fields, respectively.

\section{Modification of Generalized CT via Gauge Property of Waves of Displacement}
\indent

As the motion of an environment can make the motion of a wave
within it ineffective, it is noteworthy to mention that when a
wave of displacement moves a microstructure of medium, then the
microstructure movement can make the movement of waves of rotation
(the gradient or the momentum of waves of rotation) ineffective.
This matter means that the displacement field is capable to play
the role of the momentum of waves of rotation, similar to the
photon field that acts as the role of the momentum or a gauge for
the fermion field. However, there is no trace of such a feature in
Eqs.~\eqref{dis-eq-x} and \eqref{dis-eq}. Hence, we still need to
modify the generalized CT further in order to present such a gauge
characteristic. For this purpose, let us first check whether the
existed generalized CT without the interacting terms between the
waves of displacement and waves of rotation (i.e., without the
last terms in Eqs.~\eqref{dis-eq-x} and \eqref{dis-eq}), is
related to the theory of Yang-Mills or not. If the result being
positive, then we will need only to modify the interaction terms.

The non-interactive part of Eq. \eqref{dis-eq} with its gauge
condition is well-known to be in accord with boson fields. For the
other equation, it is also known~\cite{Klauder} that a spinor
field, say $\psi $, can be defined by a two-form field as
\begin{equation}\label{spin-tensor}
\psi\equiv\varphi_{\alpha\beta}\sigma^{\alpha\beta} \vartheta,
\end{equation}
where $\sigma^{\alpha\beta}=\frac{i}{2}\left[
\gamma^\alpha,\gamma^\beta\right] $, $\gamma^\alpha $'s are the
Dirac matrices and $\vartheta$ is an arbitrary constant nonzero
fiducial spinor for which $ \gamma^5\vartheta\neq i\vartheta$.
Multiply the transpose conjugate of $\vartheta$ to a Klein-Gordon
equation~\cite{Gross-1993} for
$\tilde{\psi}\equiv\varphi_{\tilde{\alpha}\tilde{\beta}}\sigma^{\tilde{\alpha}\tilde{\beta}}
\vartheta$, i.e.
\begin{equation}\label{Klein--Gordon}
\vartheta^\dagger\left( i\tilde{\slashed\partial}+m\right) \left(
i\tilde{\slashed\partial}-m\right)\tilde{\psi} =0,
\end{equation}
where $\slashed\partial\equiv \gamma^{\alpha}\partial_{\alpha}$.
Then, substituting for $\tilde{\psi}$ while using
relation~\cite{Klauder}
\begin{equation}\label{produ}
8\vartheta\vartheta^\dagger=2\vartheta^\dagger\gamma_\alpha
\vartheta\gamma^\alpha+\omega_{\alpha\beta}\sigma^{\alpha\beta},
\end{equation}
in the exterior algebra, it yields
\begin{equation}\label{produExterior}
\tilde{\boldsymbol\omega}\wedge*\left(
\tilde{\delta}\tilde{d}\tilde{\boldsymbol\varphi}-m^2\tilde{\boldsymbol\varphi}\right)
=0,
\end{equation}
where $\wedge$ is the wedge product, $*$ is the Hodge
operator~\cite{Strauman}, and ${\boldsymbol\omega}$ is an
arbitrary two-form defined as $\omega^{\alpha\beta}\equiv
\vartheta^\dagger\sigma^{\alpha\beta}\vartheta$. Hence, in turn,
while assuming $m\equiv\sqrt{4\alpha/\left(2\gamma+\beta\right)}$,
it gives the non-interactive part of Eq.~\eqref{dis-eq-x}.

On the other hand, if the Dirac equation for a massive
fermion~\cite{Gross-1993}, i.e.
\begin{equation}\label{dirac-eq}
\left( i\tilde{\slashed\partial}- m\right)\tilde{\psi} =0,
\end{equation}
is satisfied, then its corresponding Klein-Gordon equation, and in
turn Eq.~\eqref{produExterior}, will be held\rlap.\footnote{The
comparison between the waves of fermions and rotation has also
been considered in Refs.~\cite{Close,Burnett,Chan}. Besides, the
anticommutation nature of the spinor field \eqref{spin-tensor} has
been described in Ref.~\cite{Deymier}, which guarantees the
characteristics of the Fermi-Dirac statistics of the waves of
rotation. However, by such a comparison, it does~not mean that the
classical behavior of waves of rotation is quantized. This is just
a comparison between the only two left- and right-turning
rotations of the classical waves of rotation and the only two up-
and down-spin of the waves of fermions. In this respect, the waves
of rotation of microstructures can be visualized as the classic
states of the waves of fermions. Incidentally, in
Refs.~\cite{Deymier,Deymier-2017,Deymier-2018}, the curl of the
waves of displacement has been considered equivalent to the waves
of fermions.}\
 To check the corresponding gauge condition (i.e.,
the second one in gauges~\eqref{dis-eq-xxx}), let us employ the
trivial condition
\begin{equation}\label{triv-cond}
\vartheta^\dagger\partial_{\check{\alpha}}\partial_{\check{\beta}}\sigma^{\check{\alpha}\check{\beta}}\check{\psi}
=0
\end{equation}
that, in turn by relation \eqref{produ}, gives
\begin{equation}\label{triv-cond-1}
\check{\boldsymbol\omega}\wedge*\left( \check{d}
\check{\delta}\check{\boldsymbol\varphi}\right)  =0.
\end{equation}
Now, as ${\boldsymbol\omega}$ is arbitrary, it yields the desired
result. In addition, the rotation field has two states of right
and left, and analogously, the spin $ 1/2 $ of fermions with its
gyromagnetic ratio causes the effective amount of spin, in
reaction to a magnetic field, being $ \pm 1 $. Therefore, the
non-interacting QED equations result the non-interacting
generalized CT.

With respect to the gauge characteristic mentioned before,
regarding the necessity of modifying the generalized CT, if we
choose Eq.~\eqref{dirac-eq} to be modified as
\begin{equation}\label{12}
\left( i\tilde{\slashed\partial}-e\tilde{\slashed u}-m\right)
\tilde{\psi}=0,
\end{equation}
then consequently, Eq. \eqref{dis-eq-x} will be amended as
\begin{equation}\label{dis-eq-xy}
\left( 2\gamma+\beta\right)
\tilde{\Delta}\tilde{D}\tilde{\boldsymbol\varphi}-4\alpha\tilde{\boldsymbol\varphi}=0,
\end{equation}
where $e$ is a constant parameter depended on the characteristics
of the elastic wave of environment, $D\equiv d+ie\mathbf{u}$ and
$\Delta\equiv *D* $. Eq.~\eqref{12} is invariant under $U(1)$
gauge transformation, which indicates that the waves of
displacement play the role of momentum for the waves of rotation.
Moreover, we need to modify Eq.~\eqref{Neq1-40000} in such a way
that also remains invariant under $U(1)$. Accordingly, we choose
\begin{equation}\label{11}
2(\alpha+\mu)F^{\bar{\alpha}\bar{\beta}}{}_{,\bar{\beta}}+e\,\overline{\tilde{\psi}}\gamma^{\tilde{\alpha}}
\tilde{\psi}=0.
\end{equation}
With the above choices, Eqs. \eqref{12} and \eqref{11} can be
gained from the variation of Lagrangian
\begin{equation}\label{qed-cosserat}
L=\overline{\tilde{\psi}}\left(
i\tilde{\slashed\partial}-e\tilde{\slashed u}-m\right)
\tilde{\psi}-\frac{1}{2}(\alpha+\mu)
F^{\bar{\alpha}\bar{\beta}}F_{\bar{\alpha}\bar{\beta}},
\end{equation}
with respect to $\overline{\tilde{\psi}} $ and $u_{\bar{\alpha}}
$, where $\overline{\tilde{\psi}}$ is the adjoint of
$\tilde{\psi}$.

\section{QED Treatment}
\indent

Now, by employing the exchange of variables as
\begin{equation}\label{eq1-11-a}
\sqrt{2(\alpha+\mu)}\, u^{\alpha}\longrightarrow
u^{\alpha}\quad\qquad {\rm and}\quad\qquad
e/\sqrt{2(\alpha+\mu)}\longrightarrow e,
\end{equation}
Lagrangian (\ref{qed-cosserat}) can be rewritten as
\begin{equation}\label{eq1-11-b}
L=\overline{\tilde{\psi}}\left(
i\tilde{\slashed\partial}-e\tilde{\slashed u}-m\right)
\tilde{\psi}-\frac{1}{4}
F^{\bar{\alpha}\bar{\beta}}F_{\bar{\alpha}\bar{\beta}}.
\end{equation}

This Lagrangian is equivalent to the QED Lagrangian for photon
with zero rest-mass and fermion with mass $ m $ and negative
charge $ e $~\cite{Peskin}, except that the metrics of different
parts of it are different. However, for a specific elastic
environment via the thermodynamic characteristics, namely the one
with properties
\begin{equation}\label{eq1-14-1-1-4}
\alpha=\mu+\lambda ,\quad\qquad \varepsilon=\gamma+\beta
\quad\qquad {\rm and}\qquad\quad
\frac{\alpha+\mu}{\rho}=\frac{\gamma+\varepsilon}{\Im},
\end{equation}
this equivalence is exact. In Lagrangian (\ref{eq1-11-b}), the
displacement vector $u^{\alpha} $ is analogous with the
electromagnetic potential four-vector in electrodynamics, and the
tensor $\varphi^{\alpha\beta} $ is comparable to the fermion spin
tensor in the theory of QED. In particular, components
$\varphi^{0i} $ are analogous with the electric dipole ones and
components $\varphi^{ij} $ are comparable to the magnetic dipole
ones.

\section{Empirical Evaluation}
\indent

The behavior of light in dispersive media can be explained using
QED~\cite{Andrews-2020}. On the other hand, the behavior of
elastic waves in a granular medium can be described by an elastic
wave theory in an asymmetric
environments~\cite{ref29,Merkel-2010}. In this situation, if the
behavior of light in dispersive media being similar to the
behavior of elastic waves in granular environments, it is
plausible to conclude that such an elastic wave theory in an
asymmetric medium would also correspond to the theory of QED.

Having this in mind, let us qualitatively justify our
modifications. Accordingly, from experimental point of view, a
report has been presented~\cite{Merkel} on the empirical
evaluation of the CT. Therein, a comparison has been made between
their experimental results with the predictions of the CT and also
a discrete lattice model~\cite{Merkel-2010}. At last, there has
been shown that the CT is at least inadequate and the other model
is better than the CT. For our purpose, we have qualitatively
compared the experimental results of~\cite{Merkel} with the
presented amended CT. In this regard, we will mainly indicate that
the elastic waves behavior in granular environments is similar to
the behavior of light in dispersive media.

In Fig.~$3$ of Ref.~\cite{Merkel} (repeated here as Fig.~$1$ with
some subregions marked on it), two regions have been specified.
The first region is the low-frequency region ($0-80$ kHz) or the
L-TR region. In this region, in two subregions of less than $ 20 $
kHz and $20-50$ kHz (the subregions $1$ and $2$ in Fig.~$1$,
respectively), the intensity of wave is maximal, and the
arrival-time is almost constant with respect to the variations of
frequency (as shown by the arrows in the subregions $1$ and $2$).
Thus, the environment for the waves of displacement is transparent
and the waves of displacement have constant speed and are
massless. However, according to their figure, this environment
consists of the subregion of less than $20$ kHz (that is related
to the longitudinal wave) and the subregion $20-50$ kHz (that
corresponds to the shear wave which, due to the intensity of wave
in these two subregions, is linearly coupled to the longitudinal
wave). Now, let us compare this result with the situation
considered in
Refs.~\cite{Pniewski--2016,Jazbinsek--2019,Kabir--2020}, wherein
the motion of light in a dispersive medium has been considered.
Their results, particularly Fig.~$9$ of Ref.~\cite{Kabir--2020},
in the region of low-energies and long-wavelengths, illustrate
that the refractive index of the medium (and therefore the speed
of light) is almost constant and does~not change with respect to
changes in the light frequency.

Furthermore, in the subregion $50-80$ kHz (the subregion $3$ in
Fig.~$1$), in addition to the decrement of the wave intensity, the
wave arrival-time increases with the increment of the wave
frequency (as shown by the arrows in the subregion $3$) and thus,
the wave speed decreases. Again, let us compare this new result
with the mentioned situation in
Refs.~\cite{Pniewski--2016,Jazbinsek--2019,Kabir--2020}. Their
results, particularly Fig.~$9$ of Ref.~\cite{Kabir--2020}, in the
region of high-energies and short-wavelengths, indicate that the
refractive index of the medium (due to the induction of electric
and magnetic dipoles and in turn, increase of its permittivity and
permeability) increases. Hence, the speed of light in the
dispersive medium decreases with respect to changes in the light
frequency. In analogous, this situation means that in this
subregion, components $\varphi^{0i} $ and $\varphi^{ij} $ get
excited and the wave $ u^{\alpha} $ gradually decays into the wave
$ \varphi^{\alpha\beta} $. Such a decay is visible in Fig.~$2b$ of
Ref.~\cite{Merkel} at the end of this region. Also, this decay is
similar to decay of high-energy cosmic gamma-rays in collision
with low-energy background lights and their pair production. In
Fig.~$9$ of Ref.~\cite{Abdo}, the spectral energy distribution of
the Crab Nebula from soft to very high-energy gamma-rays has been
depicted, which is similar to Fig.~$2b$ of Ref.~\cite{Merkel} that
shows transmittance measured in the granular crystal. This point
indicates similarity between gamma-rays treatment and the
presented amended CT.

\begin{figure}[htbp]
    \includegraphics[width=16.3cm]{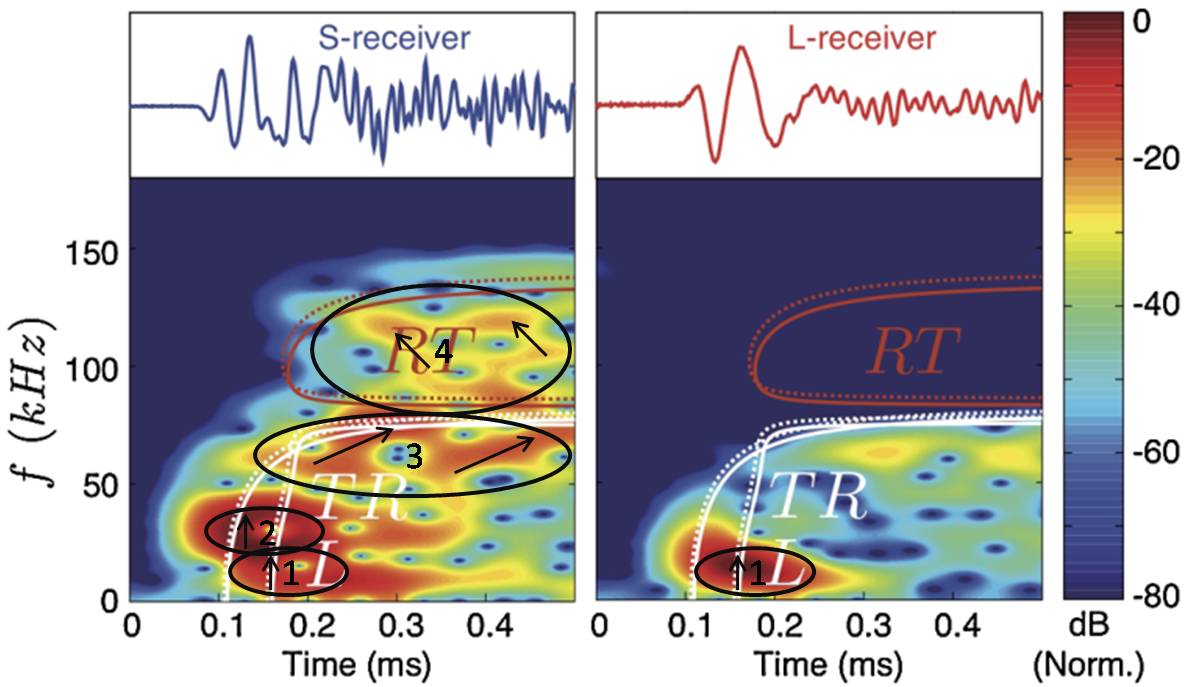}
    \caption[Caption for LOF]{\label{fig1}\small
        (colored online) This figure (the Fig~$3$ of Ref.~\cite{Merkel} with
        some marks on it) shows the arrival-time
        of received signals from a shear transducer (left) and a longitudinal
        transducer (right) in terms of frequency and intensity after transmission
        through a crystal, that we have identified thereon the variations of
        the wave velocity based on the frequency variation. }
\end{figure}

The second region in Fig.~$3$ of Ref.~\cite{Merkel} is the
high-frequency region ($80-150$ kHz, the subregion $4$ in
Fig.~$1$) or the RT-region. In this region, waves show different
behaviors compared to the first region. That is, while the
intensity of waves are very low, the wave arrival-time decreases
with the increment of frequency (as shown by the arrows in the
left-hand side of Fig.~$1$) and thus, the wave speed increases.
This issue means waves in this region are massive. Hence, the
hypothesis of the decay of massless waves $ u^{\alpha}$ into
massive waves $\varphi^{\alpha\beta} $ is confirmed, although due
to the high mass of the vibrational waves $\varphi^{i0} $, there
is no trace of this wave in the subregion $4$ in the right-hand
side of Fig.~$1$. This result can also be compared to a situation
where light with low-energies and long-wavelengths is moving in a
dispersive medium. In this case, a new phenomenon occurs after the
environmental refractive index increases sufficiently with
increment of the energy of light. This phenomenon is known as the
photovoltaic effect, which is the generation of voltage and
electric current in a material upon exposure to
light~\cite{Evans--2015,Tan--2019}. For instance, in Fig.~$3$ of
Ref.~\cite{Tan--2019}, it has been illustrated that, in the
collision of light with a wavelength in the range of $300$ to
$400$ nm, the photoelectric effect produces electric current in
TiO$_2$ nanotubes, however, at shorter and longer wavelengths,
almost no current is produced.

Also, the comparison of difference in the intensity of wave, in
the subregions $1$ and $4$ in Fig.~$1$, indicates that the wave
intensity of $\varphi^{\alpha\beta} $ is approximately the square
of the wave intensity of $ u^{\alpha}$. This result displays that,
contrary to the Cosserat and discrete models,
$\varphi^{\alpha\beta} $ wave is non-linearly coupled to $
u^{\alpha} $ wave, which is consistent with our modification of
interaction parts of the generalized CT.

Another point, which is~not a new one, is how to apply a
continuous theory to a discrete environment. In this respect, when
one quantizes a theory, one needs an energy scale to get a
renormalized theory. Such an energy scale should be proportional
to the size of grains in a granular system. If the frequency of
wave in a granular medium is close to such an energy scale, it
will be equivalent to when the wavelength approaches to the size
of system grains. In this situation, loop-corrections should be
considered and the coupling constant of U(1) gauge theory running
and being a function of energy of system~\cite{Peskin}. Such a
situation is equivalent with an effective higher-gradient theory
with increasing energy levels as it had been suggested in
Ref.~\cite{Merkel} that the CT should been combined with
higher-gradient theories like those in Ref.~\cite{Suiker}, however
therein, such terms have been manually added.

\section{Conclusions}
\indent

By considering the importance of correctly understanding the
dynamics of microstructure materials, particularly when applied in
the related technologies, while eliminating the shortcomings of
current elastic models, we have presented an asymmetric
elastodynamic model based on the QED as a $U(1)$ gauge theory with
quantum aspects. During this process, first we have noticed some
overlooked physical concepts in the Cosserat~\cite{Cosserat} and
the corresponding discrete models~\cite{Merkel-2010}, such as
disregarding the possibility of vibrational degrees of freedom for
microstructures and also the gauge property of $u^{\alpha} $ wave
for $\varphi^{\alpha\beta} $ wave. Accordingly, we have studied a
particular version of linear Cosserat elasticity, i.e. elasticity
theory which allows for microrotations of material points.

Then, we have generalized the CT in a four-dimensional form by
adding vibrational degrees of freedom for microstructures. For
this purpose, we first have shown an equivalency between, a
microstructure with rotational and vibrational degrees of freedom
in $3$-dimensions, and the same microstructure with a rotational
degree of freedom in $4$-dimensions. Also, by using some
mathematical methods and substituting the notion of multi-metrics
instead of applying several velocities for different waves (such
as longitudinal, transverse, and rotational waves), we have
expressed the wave equations in a covariant form.

Subsequently, via a gauge property of the waves of displacement
relative to waves of rotation, we have amended the existing
interaction terms of the current elastic models instead of somehow
modifying those manually. Thus, we have established a fruitful
analogy between different subfields of physics and have shown that
in asymmetric elastic environments, the elastic wave equations
have the QED structure as a $U(1)$ gauge theory. Indeed, we have
shown that the presented asymmetric elastodynamic model, based on
this gauge theory, produces solutions which exhibit similarities
with bosons and fermions in the theory of QED. That is, on macro
scales, elastic waves can possess the QED
treatment\rlap.\footnote{Also, in Ref.~\cite{PointTheory}, we have
shown a kind of symmetry that statistically implies uniformity of
physics in large and small scales.}\
 This analogy
provides a new paradigm of fermions and bosons. On the other hand,
due to the phenomenon of loop corrections and the existence of
cut-off energy in $U(1)$ gauge theories (such as QED), the
internal length scale of discontinuous materials can be comparable
to the inverse of cut-off energy. Hence in this process, in
addition of being able to take the advantage of continuum
techniques, the challenge in describing discontinuous materials
can be remedied.

Moreover, from the experimental point of view, due to the analogy
between wave intensity of $ \varphi^{\alpha\beta} $ and
$u^{\alpha} $, we have qualitatively indicated that, contrary to
the Cosserat and the corresponding discrete models, our amended
CT, as an asymmetric elastodynamic model based on a $U(1)$ gauge
theory, are consistent with the corresponding empirical
observations. In this point of view, an elastic environment
resembles an ether\footnote{For a review on the ether, see e.g.,
Ref.~\cite{FarYou} and references therein.}\
 or a dispersive medium, and the elastic waves
resemble the electromagnetic waves in that ether or dispersive
medium\rlap.\footnote{Indeed, in another work~\cite{YousefFarh2},
we have proposed an ethereal model based on a `third kind' of
quantization approach~\cite{YousefFarh5}, in which the
electromagnetic waves analogize the elastic waves in an elastic
environment.}\
 Therefore, we in fact have shown that the behavior of elastic
waves in granular media is equivalent to behavior of light in
dispersive media, which can be explained using QED, i.e. a kind of
asymmetric-elastodynamics/QED correspondence.

The importance of this correspondence, as an attempt to arrive at
an acceptable elastodynamic model, becomes more apparent when it
is applied in the engineering of modern composite materials, e.g.
the functionally graded materials (FGMs) that have had significant
impact on design and construction technology. In this regard,
discussion on the behavioral analysis of the wave propagation in
the FGMs has recently gained a wide range of research interests,
see e.g.,
Refs.~\cite{Karami--2019,Karami--2019-2,Furjan--2020,Khadimallah--2020,Asghar--2020,
Furjan--2021,Batou---2019,Bouanati--2019,Furjan---2020,Karami---2019,Bendenia---2020,Heidari---2021,Arshid---2021}.
However, the point of recent studies is that, basically, the
Coserat degree of freedom is~not usually
considered~\cite{Furjan--2020,Karimiasl,Njim}. Whereas, as
mentioned in the previous section, the effect of such a degree of
freedom can be observed at high-frequencies, and plays an
important role in the wave behavior. Thus, if one wants to study
the behavior of high-frequency waves in these materials, one will
need a genuine elastodynamic model that can accurately explain the
wave behavior in these materials, wherein the presented
correspondence with the QED might also be fruitful.

\section*{Acknowledgements}
\indent

We thank the Research Council of Shahid Beheshti University.

%

\begin{thebibliography}{1}
\bibitem{Karami--2019}B. Karami, M. Janghorban and A. Tounsi,
    ``On exact wave propagation analysis of triclinic material using three-dimensional
    bi-Helmholtz gradient plate model", {\it Struct. Eng. Mech.}\ {\bf 69} (2019), 487.
\bibitem{Karami--2019-2}B. Karami, M. Janghorban and A. Tounsi,
    ``Wave propagation of functionally graded anisotropic nanoplates resting on
    Winkler-Pasternak foundation", {\it Struct. Eng. Mech.}\ {\bf 70} (2019), 55.
\bibitem{Furjan--2020}M.S.H. Al-Furjan, M. Habibi, D. won Jung,
    S.F. Sadeghi, H. Safarpour, A. Tounsi and G. Chen, ``A computational framework for
    propagated waves in a sandwich doubly curved nanocomposite panel", {\it Eng. Comput.}\
    (2020), https ://doi.org/10.1007/s0036 6-020-01130 -8.
\bibitem{Asghar--2020}S. Asghar, M.N. Naeem, M. Hussain and
    A. Tounsi, ``Nonlocal vibration of DWCNTs based on Fl\"{u}gge shell model using wave
    propagation approach", {\it Steel Compos. Struct.}\ {\bf 34} (2020), 599.
\bibitem{Khadimallah--2020}M.A. Khadimallah, M. Hussain, K.M.
    Khedher, M.N. Naeem and A. Tounsi, ``Backward and forward rotating of FG ring support
    cylindrical shells", {\it Steel Compos. Struct.}\ {\bf 37} (2020), 137.
\bibitem{Furjan--2021}M.S.H. Al-Furjan, M.A. Oyarhossein, M.
    Habibi, H. Safarpour, D. won Jung and A. Tounsi, ``On the wave propagation of the multi-scale
    hybrid nanocomposite doubly curved viscoelastic panel", {\it Compos. Struct.}\ {\bf 255} (2021), 112947.
\bibitem{Batou---2019}B. Batou, M. Nebab, R. Bennai, H.A.
    Atmane, A. Tounsi and M. Bouremana, ``Wave dispersion properties in imperfect sigmoid plates
    using various HSDTs", {\it Steel Compos. Struct.}\ {\bf 33} (2019), 699.
\bibitem{Bouanati--2019}S. Bouanati, K.H. Benrahou, H.A.
    Atmane, S.A. Yahia, F. Bernard, A. Tounsi and G. Chen, ``Investigation of wave propagation in
    anisotropic plates via quasi 3D HSDT", {\it Geomech. Eng.}\ {\bf 18} (2019), 85.
\bibitem{Karami---2019}B. Karami, M. Janghorban and A.
    Tounsi, ``Vibration response and wave propagation in FG plates resting on elastic foundations
    using HSDT", {\it Struct. Eng. Mech.}\ {\bf 69} (2019), 487.
\bibitem{Furjan---2020}M.S.H. Al-Furjan, M. Habibi, J. Ni,
    D. won Jung and A. Tounsi, ``Frequency simulation of viscoelastic multi-phase reinforced fully
    symmetric systems", {\it Eng. Comput.}\ (2020), https://doi.org/10.1007/s00366-020-01200-x.
\bibitem{Bendenia---2020}N. Bendenia, {\it et al.},
    ``Deflections, stresses and free vibration studies of FG-CNT reinforced sandwich plates
    resting on Pasternak elastic foundation", {\it Comput. Concr.}\ {\bf 26} (2020), 213.
\bibitem{Heidari---2021}F. Heidari, K. Taheri, M. Sheybani,
    M. Janghorban and A. Tounsi, ``On the mechanics of nanocomposites reinforced by
    wavy/defected/aggregated nanotubes", {\it Steel Compos. Struct.}\ {\bf 38} (2021), 533.
\bibitem{Arshid---2021}E. Arshid, M. Khorasani, Z.
    Soleimani-Javid, S. Amir and A. Tounsi, ``Porosity-dependent vibration analysis of FG
    microplates embedded by polymeric nanocomposite patches considering hygrothermal effect
    via an innovative plate theory", {\it Eng. Comput.}\ (2021), https://doi.org/10.1007/s00366-021-01382-y.
\bibitem{Riahi}A. Riahi and J.H. Curran, ``Full 3D finite element Cosserat formulation with application
         in layered structures", {\it Appl. Math. Model.}\ {\bf 33} (2009), 3450.
\bibitem{Kelvin}W. Thomson, ``On the propagation of laminar motion
         through a turbulently moving inviscid liquid", {\it Philoso.
         Magaz. Ser.}\ {\bf 5} (1887), 342.
\bibitem{Dmitriyev92}V.P. Dmitriyev, ``The elastic model of physical
         vacuum'', {\it Mech. Solids (N.Y.)}\ {\bf 26} (1992), 60.
\bibitem{Dmitriyev98}V.P. Dmitriyev, ``Towards an exact mechanical analogy of particles and fields",
         {\it Nuov. Cim.} {\bf 111A} (1998), 501.
\bibitem{Dmitriyev01}V.P. Dmitriyev, ``Elasticity and electromagnetism (the Coulomb gauge)",
         {\it Electromag. Phenomena} {\bf 2} (2001), 474.
\bibitem{Yoshida}S. Yoshida, ``Interpretation of mesomechanical behaviors of plastic deformation
         based on analogy to Maxwell electromagnetic theory", {\it Phys. Mesomech.} {\bf 4} (2001), 29.
\bibitem{Cosserat}E. Cosserat and F. Cosserat, ``\textit{Th\'{e}orie des
         Corps D\'{e}formables} (Theory of Deformable Bodies)'', (Herman, Paris, 1909).
\bibitem{ref28}E.B. Wilson, ``An advance in theoretical mechanics", \textit{Bull.
         Am. Math. Soc.} \textbf{19} (1913), 242.
\bibitem{Toupin}R.A. Toupin, ``Theories of elasticity with couple-stress'', {\it Arch.
         Rational Mech. Anal.} {\bf 17} (1964), 85.
\bibitem{Vardoulakis}I. Vardoulakis, ``\textit{Cosserat Continuum Mechanics With Applications to Granular
         Media}'', (Springer, Switzerland, 2019).
\bibitem{ref29}W. Nowacki, ``The linear theory of micropolar elasticity'', In:
         ``{\it Micropolar Elasticity}'', Symposium organized by the department of mechanics
         of solids, June 1972, Edited by: W. Nowacki and W. Olszak, (Springer, Vienna,
         1974), pp. 1-43.
\bibitem{Stanzial}D. Stanzial, D. Bonsi and G. Schiffrer, ``Four-dimensional
         treatment of linear acoustic fields and radiation pressure", \textit{Act.
         Acust. Un. Acustica} \textbf{88} (2002), 213.
\bibitem{Kinsler-1999}L.E. Kinsler, A.R. Frey,  A.B. Coppens and J.V. Sanders,
         ``\textit{Fundamentals of Acoustics}", (Wiley, New York, 1999).
\bibitem{Harris2004}J.G. Harris, ``{\it Linear Elastic Waves}'', (Cambridge
         University Press, Cambridge, 2004).
\bibitem{Strauman}N. Straumann, ``\textit{General Relativity with Applications to Astrophysics}",
         (Springer, Berlin, 2004).
\bibitem{Klauder}J.R. Klauder, ``Linear representation of spinor fields by antisymmetric tensors",
         \textit{J. Math. Phys.} \textbf{5} (1964), 1204.
\bibitem{Close}R.A. Close, ``Exact description of rotational waves in an elastic solid'',
         {\it Adv. Appl. Clifford Algeb.}\ {\bf 21} (2011), 273.
\bibitem{Gross-1993}F. Gross, ``\textit{Relativistic Quantum Mechanics and Field Theory}",
         (Wiley, New York, 1993).
\bibitem{Burnett}J. Burnett and D. Vassiliev ``Modeling the electron with
         Cosserat elasticity'', {\it Mathematika}\ {\bf 58} (2012), 349.
\bibitem{Chan}C.T. Chan, Z.H. Hang and X. Huang, ``Dirac dispersion in two-dimensional photonic crystals'',
         {\it Adv. OptoElectron}\ {\bf 2012} (2012), 313984.
\bibitem{Deymier}P.A. Deymier, K. Runge, N. Swinteck and K. Muralidharan, ``Torsional topology and
         fermion-like behavior of elastic waves in phononic structures'',
         {\it Comptes Rendus M\'{e}canique}\ {\bf 343} (2015), 700.
\bibitem{Deymier-2017}P.A. Deymier and K. Runge, ``\textit{Sound Topology, Duality,
         Coherence, and Wave-Mixing: An Introduction to the Emerging New Science of
         Sound}", (Springer, Switzerland, 2017).
\bibitem{Deymier-2018}P.A. Deymier, K. Runge, P. Lucas and J.O. Vasseur, ``Spacetime representation of
         topological phononics", \textit{New J. Phys.} \textbf{20} (2018), 053005.
\bibitem{Peskin}M.E. Peskin and D. Schroeder, ``\textit{An Introduction to Quantum
         Field Theory}", (Westview, New York, 1995).
\bibitem{Andrews-2020}D.L. Andrews, D.S. Bradshaw, K.A. Forbes and A. Salam, ``Quantum electrodynamics in modern
         optics and photonics: Tutorial", \textit{J. Opt. Soc. Am. B} \textbf{37} (2020), 1153.
\bibitem{Merkel-2010}A. Merkel, V. Tournat and V. Gusev, ``Dispersion of elastic waves in three-dimensional
         noncohesive granular phononic crystals: Properties of rotational modes", \textit{Phys. Rev. E} \textbf{82}
         (2010), 031305.
\bibitem{Merkel}A. Merkel, V. Tournat and V. Gusev, ``Experimental evidence of rotational elastic waves in granular phononic
         crystals", \textit{Phys. Rev. Lett.} \textbf{107} (2011), 225502.
\bibitem{Pniewski--2016}J. Pniewski, {\it et al.}, ``Dispersion engineering in nonlinear soft glass
         photonic crystal fibers infiltrated with liquids", \textit{Appl. Opt.} \textbf{55} (2016), 5033.
\bibitem{Jazbinsek--2019}M. Jazbinsek, U. Puc, A. Abina and A. Zidansek, ``Organic crystals for THz photonics",
         \textit{Appl. Sci.} \textbf{9} (2019), 882.
\bibitem{Kabir--2020}I.I. Kabir, {\it et al.}, ``Contamination of TiO$_2$ thin films spin coated
         on borosilicate and rutile substrates", \textit{J. Mater. Sci.} \textbf{55} (2020), 3774.
\bibitem{Abdo}A.A. Abdo, {\it et al.}, ``Fermi large area telescope observations of the Crab pulsar and nebula",
         \textit{Astrophys. J.} \textbf{708} (2010), 1254.
\bibitem{Evans--2015}C.M. Evans, K. Krynski, Z. Streeter and G.L. Findley, ``Energy of the quasi-free electron
         in H$_2$, D$_2$ and O$_2$: Probing intermolecular potentials within the local Wigner-Seitz model",
         \textit{J. Chem. Phys.} \textbf{143} (2015), 224303.
\bibitem{Tan--2019}Y. Tan, Y. Xu, K. Liang and S. Zhang, ``Study on preparation and photocurrent response properties
         of In/In$_2$O$_3$/TiO$_2$ nanotubes arrays compound heterojunction semiconductor",
         \textit{IOP Conf. Ser.: Mater. Sci. Eng.} \textbf{472} (2019), 012006.
\bibitem{Suiker} A.S.J. Suiker, R. de Borst and C.S. Chang, ``Micro-mechanical modeling of granular material",
         \textit{Acta Mech.} \textbf{149} (2001), 161.
\bibitem{PointTheory}M. Yousefian and M. Farhoudi, ``Towards real amplituhedron via one-dimensional
         theory", {\it arXiv: 1910.05548}.
\bibitem{FarYou}M. Farhoudi and M. Yousefian, ``Ether and relativity", {\it Int. J. Theor. Phys.}
         \textbf{55} (2016), 2436.
\bibitem{YousefFarh2}M. Yousefian and M. Farhoudi, ``Fine structure constant variation and ultra
         high energy cosmic rays via a novel ethereal model", work in progress.
\bibitem{YousefFarh5}M. Yousefian and M. Farhoudi, ``Gravity as a `third' quantization gauge theory with Minkowski metric,
         non-zero torsion and curvature", {\it arXiv: 2104.05768}.
\bibitem{Karimiasl}M. Karimiasl, F. Ebrahimi and M. Vinyas, ``Nonlinear vibration analysis of multiscale doubly
         curved piezoelectric composite shell in hygrothermal environment",
         \textit{ J. Intell. Mater. Syst. Struct.} \textbf{30} (2019), 1594.
\bibitem{Njim}E.K. Njim, M. Al-Waily and S.H. Bakhy, ``A critical review
         of recent research of free vibration and stability of functionally
         graded materials of sandwich plate", \textit{IOP Conf. Ser.:
         Mater. Sci. Eng.} \textbf{1094} (2021), 012081.
%
\end{thebibliography}
\end{document}